\documentclass[%
 aip,
 amsmath,amssymb,
 reprint,%
]{revtex4-2}

\usepackage{graphicx}
\usepackage{dcolumn}
\usepackage{bm}

\usepackage[utf8]{inputenc}
\usepackage[T1]{fontenc}
\usepackage{mathptmx}
\usepackage{etoolbox}

\makeatletter
\def\@email#1#2{%
 \endgroup
 \patchcmd{\titleblock@produce}
  {\frontmatter@RRAPformat}
  {\frontmatter@RRAPformat{\produce@RRAP{*#1\href{mailto:#2}{#2}}}\frontmatter@RRAPformat}
  {}{}
}%
\makeatother
\begin{document}

\preprint{AIP/123-QED}

\title[GWPE]{Generalized Weighted Permutation Entropy}

\author{Darko Stosic}
\author{Dusan Stosic}
\affiliation{Centro de Inform\'atica, Universidade Federal de Pernambuco, Av. Luiz Freire s/n, 50670-901, Recife, PE, Brazil}
\author{Tatijana Stosic}
\author{Borko Stosic}
 \altaffiliation{Author to whom correspondence should be addressed:\\ borko.stosic@ufrpe.br}
\affiliation{
Departamento de Estat\' \i stica e Inform\' atica, 
Universidade Federal Rural de Pernambuco,\\
Rua Dom Manoel de Medeiros s/n, Dois Irm\~ aos,
52171-900 Recife-PE, Brazil
}

\date{\today}

\begin{abstract}
A novel heuristic approach is proposed here for time series data analysis, dubbed Generalized weighted permutation entropy, which amalgamates and generalizes beyond their original scope two well established data analysis methods: Permutation entropy, and Weighted permutation entropy. The method introduces a scaling parameter to discern the disorder and complexity of ordinal patterns with small and large fluctuations. Using this scaling parameter, the complexity-entropy causality plane is generalized to the complexity-entropy-scale causality box. Simulations conducted on synthetic series generated by stochastic, chaotic, and random processes, as well as real world data, are shown to produce unique signatures in this three dimensional representation.
\end{abstract}

\maketitle

\begin{quotation}
Permutation entropy (PE) was introduced 
two decades ago 
\cite{bandt2002permutation} to quantify disorder of a time series in terms of local ordinal patterns (rank vectors of observed value segments), and has subsequently been successfully used in a large number of studies. On the other hand, while PE captures local ordering of the series within segments of a given size, it disregards the magnitude of the fluctuations within these segments. To address this issue Weighted permutation entropy (WPE) was introduced 
a decade ago 
\cite{fadlallah2013weighted} to account for the variance in magnitudes observed in each pattern, and has been shown to outperform the original PE in a number of studies. In this work we amalgamate and generalize these two methods through a novel continuous scaling parameter, so that both small and large fluctuations are emphasized. 

\end{quotation}

\section{Introduction}
The conventional statistical methods have been traditionally the principal means of ``making sense'' of observational data when there is lack of {\it a priori} knowledge of the underlying mechanism of the data generating process, but the more recent and less conventional methods are becoming ever more employed in the current body of scientific literature. One such method is Permutation entropy (PE), introduced by Bandt and Pompe \cite{bandt2002permutation} to quantify disorder in a time series by taking into account local ordering of values. The PE method has been widely applied (both in its original form and in its variants) in physiology \cite{zeng2018characterizing}, engineering \cite{gao2017multi}, geophysics \cite{consolini2014permutation}, climatology \cite{barreiro2011inferring}, hydrology \cite{mihailovic2014complexity}, and finances \cite{zunino2009forbidden}. A shortcoming of this method is that patterns with identical orderings give the same contribution to entropy regardless of the magnitudes in their original values and corresponding fluctuations. Fadlallah et al. \cite{fadlallah2013weighted} address this issue through a modification in the probability distribution that accounts for the variance in magnitudes observed in each pattern. In this work we further generalize the definition of the probability distribution to discern the disorder and complexity of ordinal patterns with varying degrees of fluctuations, both small and large, through a novel     scaling parameter. Our simulations across stochastic, chaotic, and random processes show that the disorder-complexity varies significantly across scales.

\section{Permutation entropy}
The original formulation of the PE method works as follows \cite{bandt2002permutation}. For a given dimension $d>1$, from the time series $x_t, t=1,…,T$, overlapping segments  $X_s=(x_s,x_{s+1},\dots,x_{s+d-1})$, $s=1,\dots,S$ of length $d$ are extracted, where $S=T-(d-1)$. Within each segment the values are sorted in increasing order $x_{s+r_0}\leq x_{s+r_1}\leq \dots \leq x_{s+r_{d-1}}$, and the corresponding vector of indices $v_s\equiv(r_0,r_1,\dots,r_{d-1})$ is taken to represent symbolically the original segment. Each vector corresponds to a particular permutation $\pi_i$ , $i=1,\dots,d!$ of the set of integers ${0,1,\dots,d-1}$, and the relative frequencies of permutations $\pi_i$ represent the empirical probability distribution $\Pi\equiv\{p(\pi_i ),i=1,\dots,d!\}$, through which permutation entropy is defined as 
\begin{equation}
H(d)=-\sum_{i=1}^{d!} p(\pi_i ) \log{p(\pi_i )}	\;\; .
\label{eq1}
\end{equation}
Permutation entropy $H(d)$ assumes values in the range $0\leq H(d) \leq \log d!$, where the lower bound corresponds to strictly increasing or decreasing series when only a single permutation $\pi_i$ is observed with probability $p(\pi_i)=1$, and the upper bound corresponds to a random series where all the d! possible permutations are observed with the same probability $p(\pi_i )=1/d!$. In order to guarantee good statistics, the typical convention \cite{riedl2013practical} is to use the maximum $d$ that satisfies the condition $T>5d!$. 

\section{Weighted Permutation Entropy}
While the method described above captures local ordering of a series within segments of size $d$, it fails to take into account the magnitude of fluctuations. Fadlallah et al. \cite{fadlallah2013weighted} addressed this with Weighted permutation entropy (WPE), by taking into account the variance observed in each segment. Rather than using relative frequencies of patterns for the empirical probability distribution $p(\pi)$, the probability of pattern $\pi_i$ is defined as \cite{fadlallah2013weighted}
\begin{equation}
p(\pi_i)=\frac{\sum_{s,\pi_s=\pi_i} w_s}{\sum_{s} w_s} \;\; ,
\label {eq2}
\end{equation}
where $w_s$ is the variance of values observed in segment $s$, given by $w_s=1/d \sum_{i=0}^{d-1}(x_{s+i}-\left<x_s \right>)^2$  and $\left<x_s\right>=1/d \sum_{i=0}^{d-1}x_{s+i}$  is the average. Weighted permutation entropy is then obtained by inserting (\ref{eq2}) into (\ref{eq1}). Since the WPE definition is sensitive to the magnitude of fluctuations in the data, it has been shown to work better than the original PE definition in a number of situations
\cite{deng2015complexity,yin2014weighted,chen2015experimental,bian2016weighted,gan2018rolling}.

\section{Generalized Weighted Permutation Entropy}
The principal contribution of the current work is to further generalize the definition of the pattern probability distribution in order to discern between the effects of both small and large fluctuations. 
This is accomplished through the novel pattern probability definition
\begin{equation}
p(\pi_i,q)=\frac{\sum_{s,\pi_s=\pi_i}^{'} w_s^{q/2}}{\sum_{s}^{'} w_s^{q/2}} \;\; ,
\label {eq3}
\end{equation}
where $q$ is a continuous scaling parameter $-\infty<q<\infty$. While $q<0$ enhances the contribution of small fluctuations on $p(\pi_i,q)$, $q>0$ enhances those of large fluctuations. Of note, if the variance $w_s$ of a given segment is strictly zero, then it does not contribute to the expression  (\ref{eq2}), but would lead to divergence in (\ref{eq3}) for negative values of $q$. Therefore, we adopt the convention that zero $w_s$ values are omitted from the summation, as denoted by the prime symbol in the sums of (\ref{eq3}). For each value of $q$, values $p(\pi_i,q)$ are then plugged into (\ref{eq1}) to define the generalized weighted permutation entropy (GWPE).
Note that PE and WPE represent special cases of GWPE for $q=0$  and $q=2$, respectively, while $q=1$ corresponds to weighting the observed patterns with the standard deviation of their values. 

\section{Complexity-Entropy-Scale Causality Box}
Lamberti et al. \cite{lamberti2004intensive} introduced the complexity-entropy causality plane (CECP) to simultaneously quantify both the information content and structural complexity of a time series. CECP was shown useful in distinguishing between stochastic noise and deterministic chaotic behavior \cite{rosso2007distinguishing}, leading to many applications in analyzing data from physiology \cite{legnani2018analysis}, physics 
\cite{weck2015permutation,maggs2013permutation},
oceanography \cite{siddagangaiah2016complexity}, 
ecology \cite{sippel2016diagnosing}, 
hydrology \cite{stosic2016investigating} 
and finances \cite{bariviera2018analysis}. 

In its original representation \cite{rosso2007distinguishing}, the horizontal axis of CECP corresponds to PE, and the vertical axis is a statistical complexity measure
\begin{equation}
C[P]=\frac{J[P,U]}{J_{max}}H_s[P] \quad ,
\label {eq4}
\end{equation}
where $P\equiv\left\{p(\pi_i ),i=1,…,d!\right\}$ is the Bandt-Pompe probability distribution and $H_s[P]=H[P]/\log{d!}$ is the normalized PE. $J[P,U]$ is the Jensen-Shannon divergence
\begin{equation}
J[P,U]=\left\{H\left[\frac{P+U}{2}\right]-\frac{H[P]}{2}-\frac{H[U]}{2}\right\}  \quad ,
\label {eq5}
\end{equation}
which quantifies the distance of $P$ from the uniform distribution $U$, and $J_{max}$ is the maximum possible value of $J[P,U]$, obtained when one of the components of $P$ is equal to unity, while all the others are equal to zero
\begin{equation}
J_{max}=-\frac{1}{2}\left[\frac{d!+1}{d!}\log{(d!+1)}-2\log{2d!}+\log{d!}\right]
\label {eq6}
\end{equation}
The definition of statistical complexity (\ref{eq4}) guarantees that both monotonically increasing and decreasing series ($H_s[P]=0$) and completely random series ($J[P,U]=0$) have zero complexity. For each given value of the normalized PE, $H_s\in[0,1]$, there is a range of possible values of complexity, $C_{min}\leq C \leq C_{max}$, which gives the lower and upper envelopes in CECP \cite{martin2006generalized}. 

Generalizing this representation to GWPE is straightforward by adopting (\ref{eq3}) for the distribution $P\equiv{p(\pi_i ),i=1,…,d!}$ to be plugged into equations  (\ref{eq4}), (\ref{eq5}) and (\ref{eq6}). In this scenario, the structural complexity values quantify the distance from the uniform distribution in terms of both ordering and fluctuations (small or large) on the scale defined by the parameter $q$. As there are now three variables (entropy, complexity, and the magnification scale parameter $q$), we term this representation the complexity-entropy-scale causality box (CESCB).

\section{Simulations}
In what follows the generalized weighted permutation entropy is tested on a range of data: artificial time series generated from stochastic processes (fractional Brownian motion with Hust exponent of $H=0.1$, $H=0.5$ and $H=0.9$), chaotic processes (Henon map, Logistic map and Skew tent map, with same parameters as in \cite{rosso2007distinguishing}), random processes (Gaussian and Uniform distribution using Marsaglia's MWC random number generator \cite{marsaglia2003random}), and real time series from ECG signals with ventricular bigeminy coded  ``aami31a'', made available 
for testing devices that monitor the electrocardiogram \cite{association2002american}. 
All the results were calculated henceforth with $d=6$, for compatibility and comparison with \cite{rosso2007distinguishing}. The artificial series are all of length $2^{15}=32768$, while the aami31a ECG signal has 43081 data points ($60$ seconds sampled at 720Hz).

For comparisons, Tab.~\ref{tab1} presents the permutation entropy (PE) and complexity (PEC), as well as the weighted permutation entropy (WPE) and corresponding complexity (WPEC). The PE and PEC results presented in Tab.~\ref{tab1} agree with those of \cite{rosso2007distinguishing}. Interestingly, WPE exhibits significantly different values from PE for fBm signals with $H=0.5$ and $H=0.9$, and for the aami31a ECG signal, indicating that the magnitude of fluctuations play an important role.

\begin{table}[ht]
\caption{Permutation entropy (PE) and complexity (PEC), as well as Weighted permutation entropy (WPE) and complexity (WPEC), for the nine considered series.}
\centering
\begin{tabular}{lcrcrcrcrc}
\hline
\hline
&&PE&&PEC&&WPE&&WPEC\\
\hline
fBm 0.1&&0.9925&&0.0173&&0.9849&&0.0339\\
fBm 0.5&&0.8945&&0.1831&&0.7773&&0.3014\\
fBm 0.9&&0.4860&&0.3042&&0.1803&&0.1632\\
Henon&&0.5554&&0.4587&&0.5392&&0.4492\\
Logistic&&0.6291&&0.4842&&0.6159&&0.4779\\
Skew &&0.4743&&0.3986&&0.5419&&0.4442\\
Gaussian&&0.9982&&0.0042&&0.9975&&0.0060\\
Uniform&&0.9983&&0.0042&&0.9978&&0.0052\\
aami31a&&0.7525&&0.3358&&0.3531&&0.2580\\
\hline
\hline
\end{tabular}
\label{tab1}
\end{table}

\begin{figure}[h]
\begin{tabular}{cc}
\includegraphics[width=.5\linewidth]{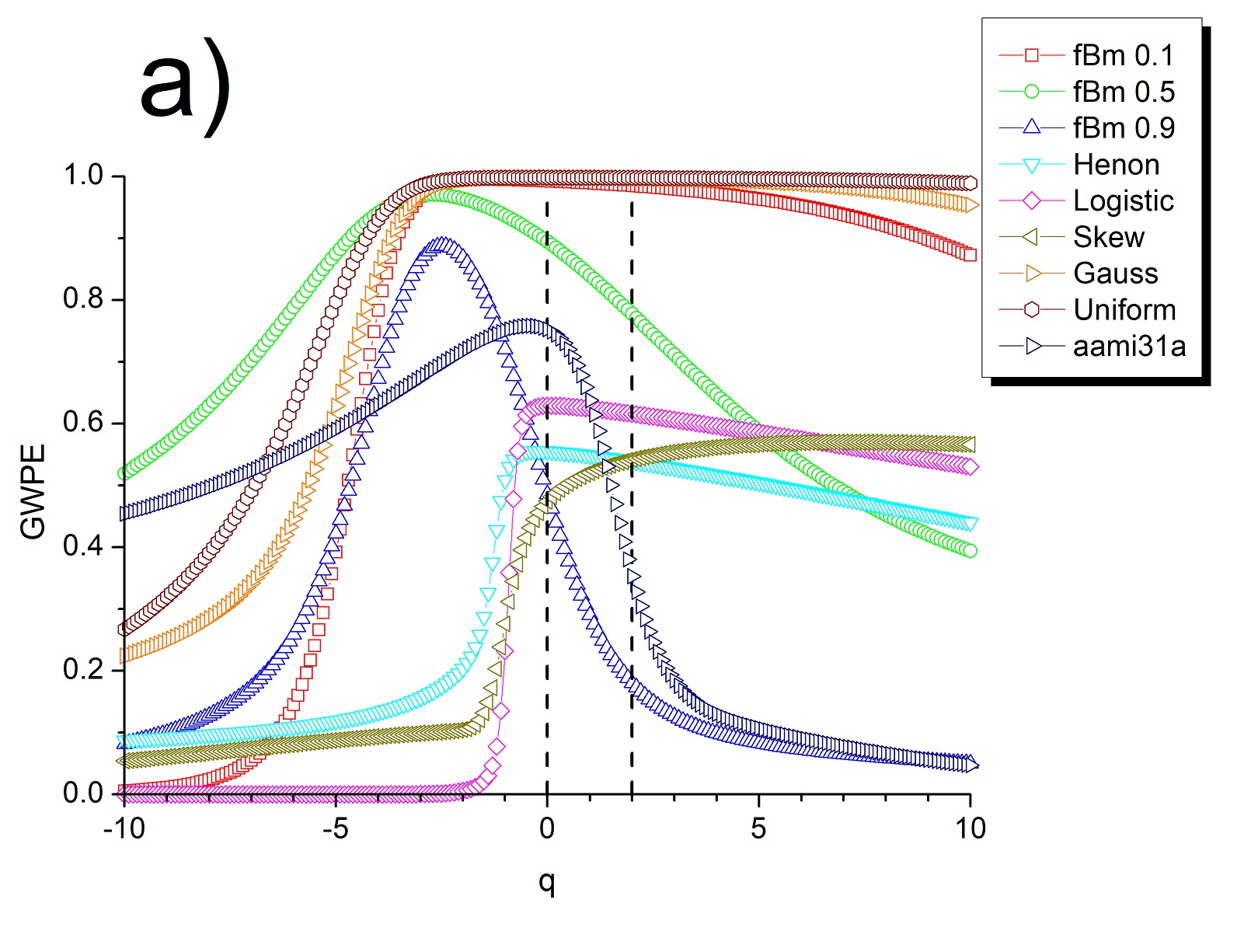} & \includegraphics[width=.5\linewidth]{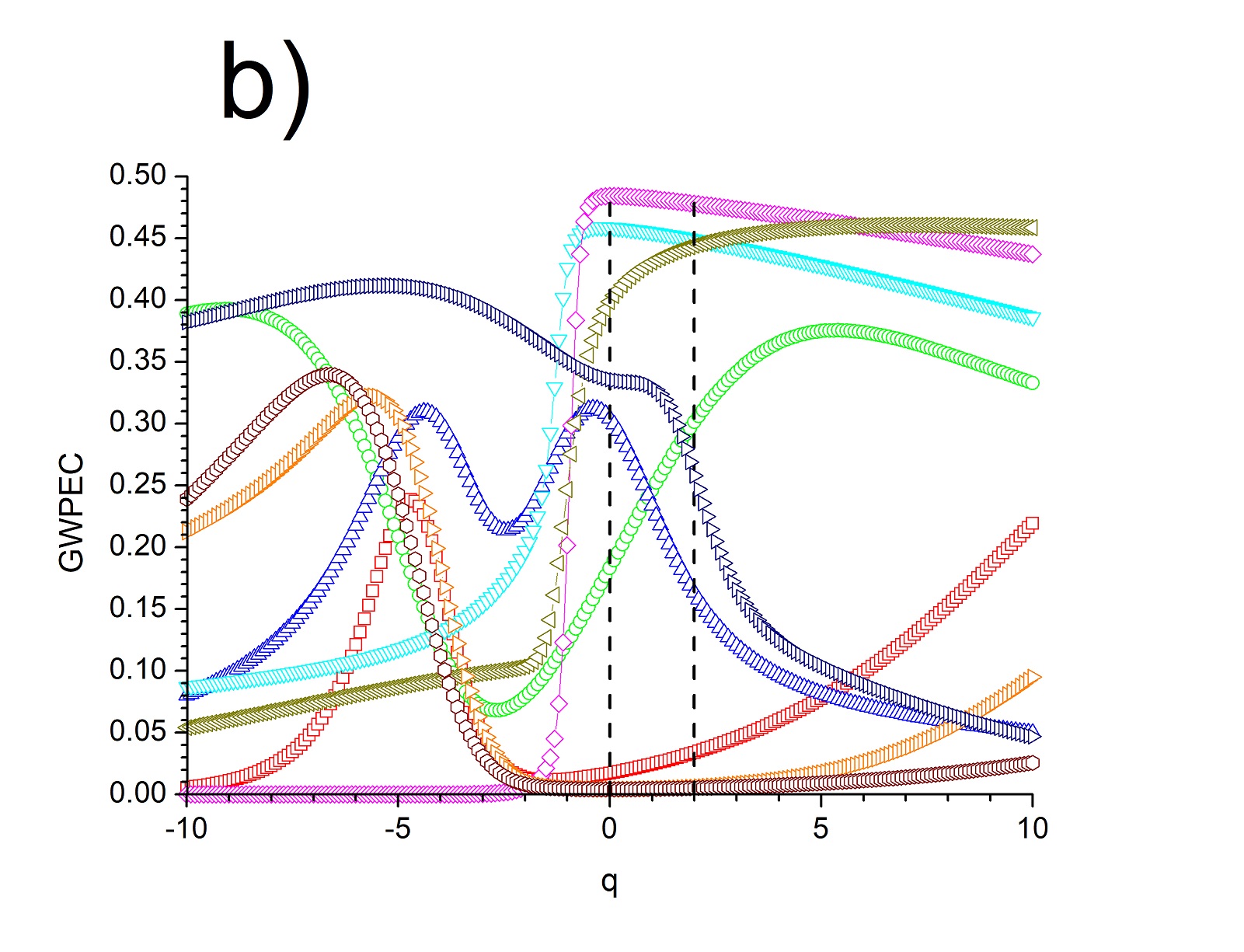}\\
\end{tabular}
\caption{a) Generalized weighted permutation entropy and b) complexity, as a function of the magnification parameter $q$. The vertical dashed lines correspond to PE and WPE.}
\label{fig1}
\end{figure}

In Fig.~\ref{fig1} the results for generalized weighted permutation entropy and corresponding complexity are displayed for values of the magnification parameter $-10\leq q\leq 10$, where a clear difference in the behavior of the curves for different signals can be observed. 
In particular, the fBm signals with $H=0.5$ and $H=0.9$, and the aami31a ECG signal demonstrate the fastest decay of entropy in the positive range of $q$, and two peaks of complexity. 

In the negative $q$ range GWPE falls below unity even for the random Gaussian and uniform signals. Namely, the variance for the normal variable samples of size $d$ follows the chi-squared distribution with $d-1$ degrees of freedom, $\chi_{d-1}^2$, and in the current case of $d=6$ it can be easily verified that the probability of attaining one tenth of the mean variance is only 0.00788. For $q=-10$ these low variance segments contribute five orders of magnitude more than the mean variance sequences, and as in the example above we are dealing with $2^{15}$ values, comprising $2^{15}-(6-1)=32763$ sequences of length $d=6$, only 258 of these are expected to yield contribution of five orders of magnitude larger than the mean in (\ref{eq3}). As there are $6!=720$ possible patterns they cannot be equally represented in the sum of (\ref{eq1}), and the entropy becomes lower. Similar reasoning holds for the uniform distribution sample, where the distribution of the variance is not known analytically \cite{weissman2017sum} for $d=6$,
but the probability of low variance word segments can be easily verified numerically.
In this sense, the fBm $H=0.5$ and the aami31a signals may be considered more ``disordered'' than the synthetic Gaussian and the uniform random number series.

\begin{figure}[h]
\begin{tabular}{ccc}
\includegraphics[width=.33\linewidth]{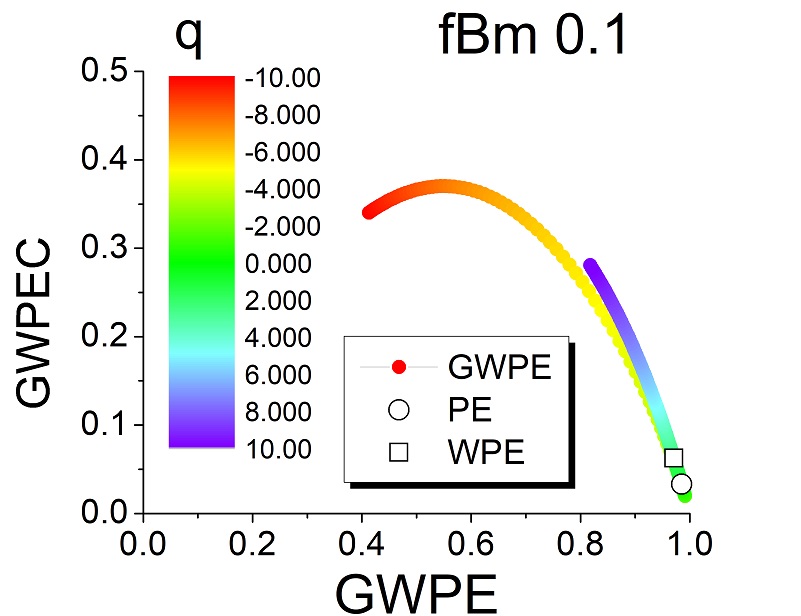} & 
\includegraphics[width=.33\linewidth]{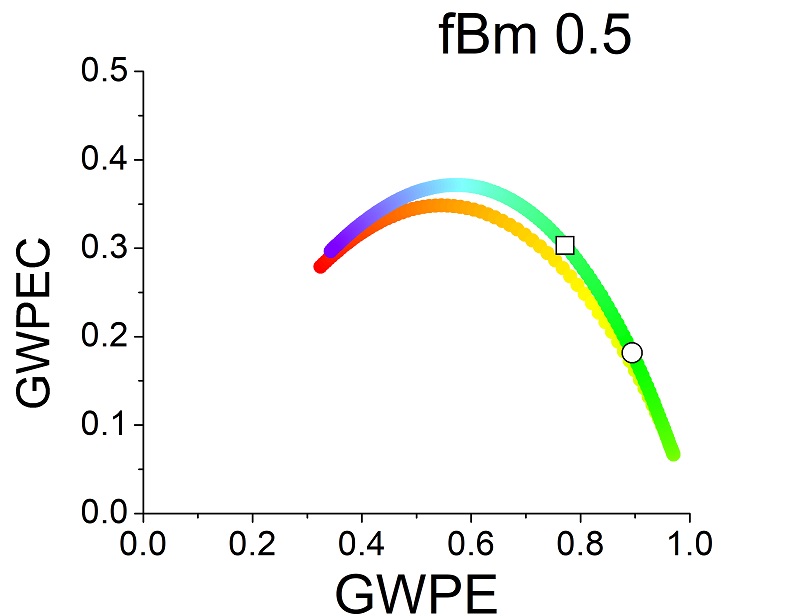} &
\includegraphics[width=.33\linewidth]{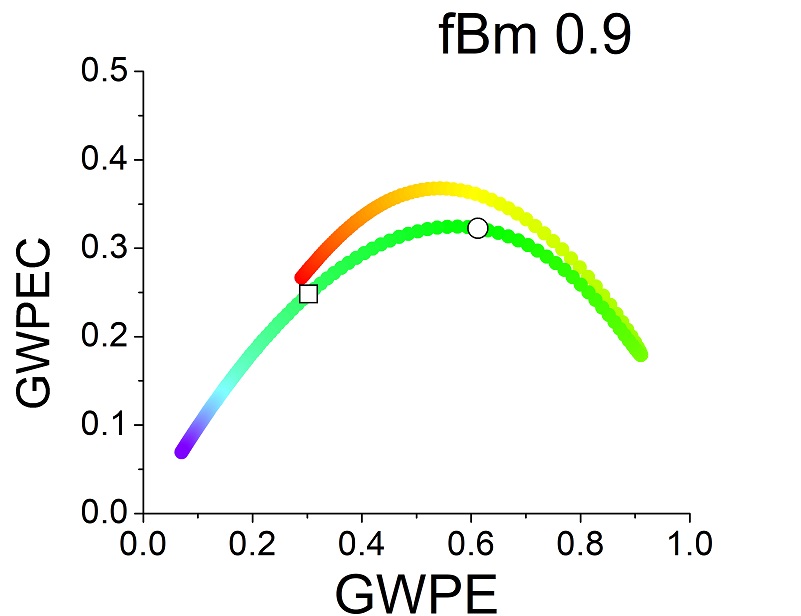}\\
\includegraphics[width=.33\linewidth]{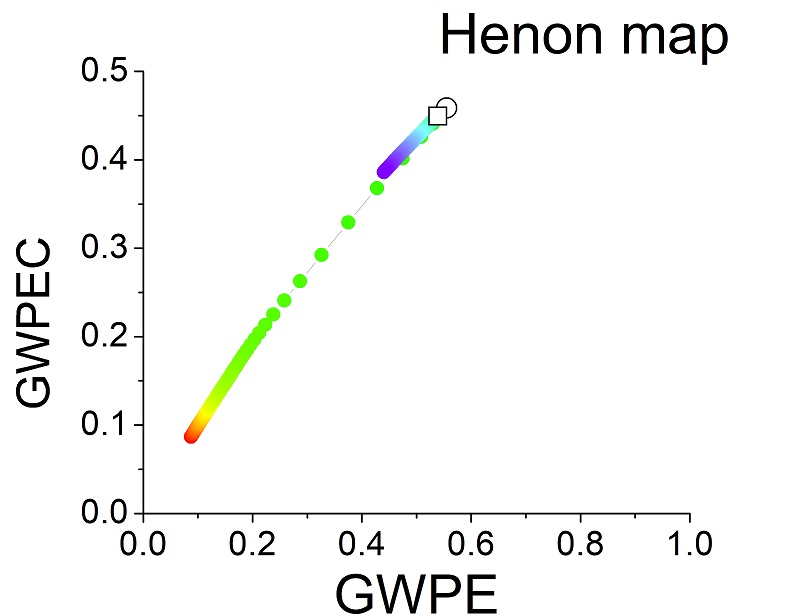} & 
\includegraphics[width=.33\linewidth]{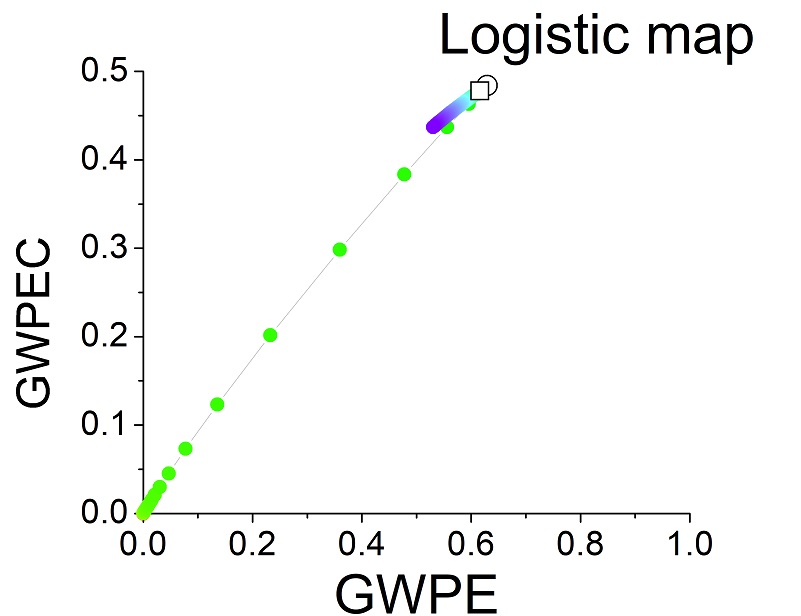} &
\includegraphics[width=.33\linewidth]{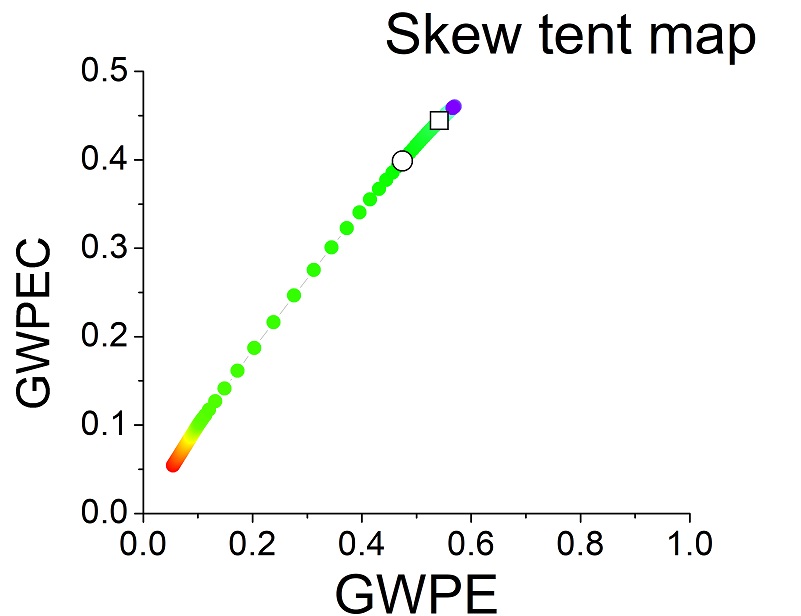}\\
\includegraphics[width=.33\linewidth]{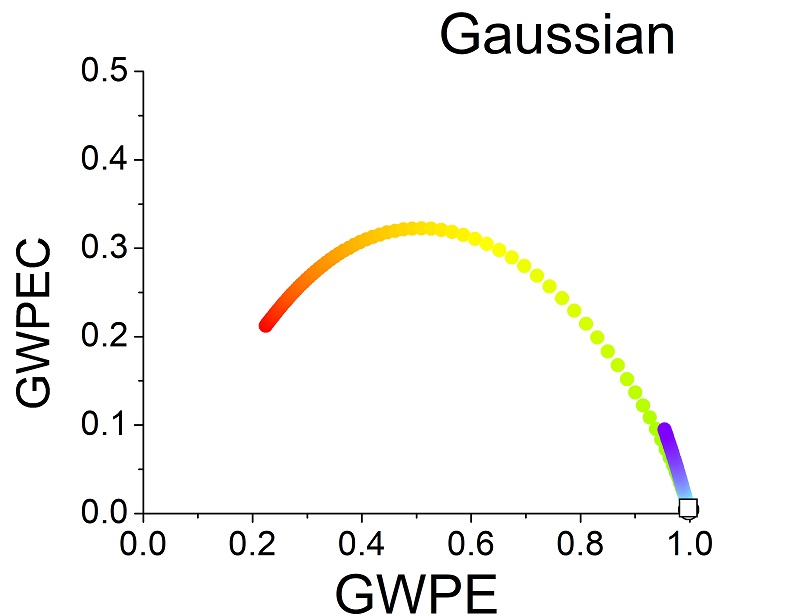} & 
\includegraphics[width=.33\linewidth]{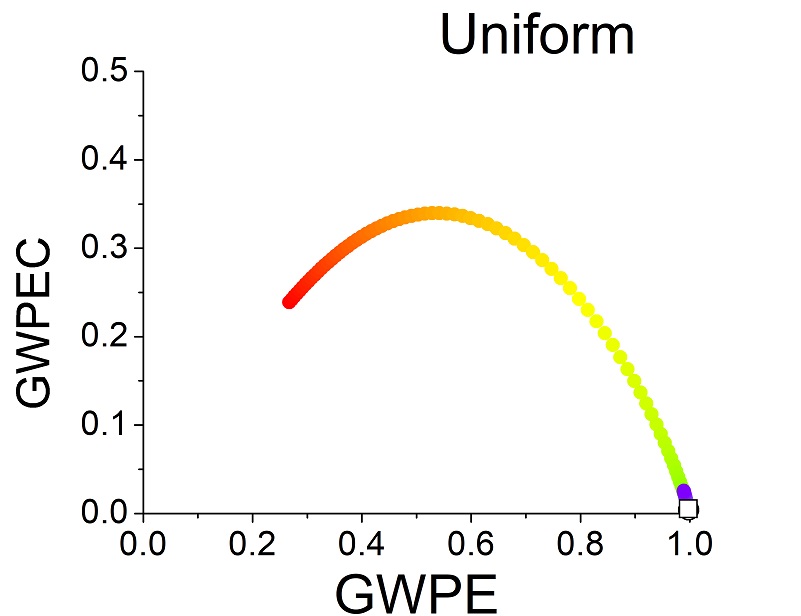} &
\includegraphics[width=.33\linewidth]{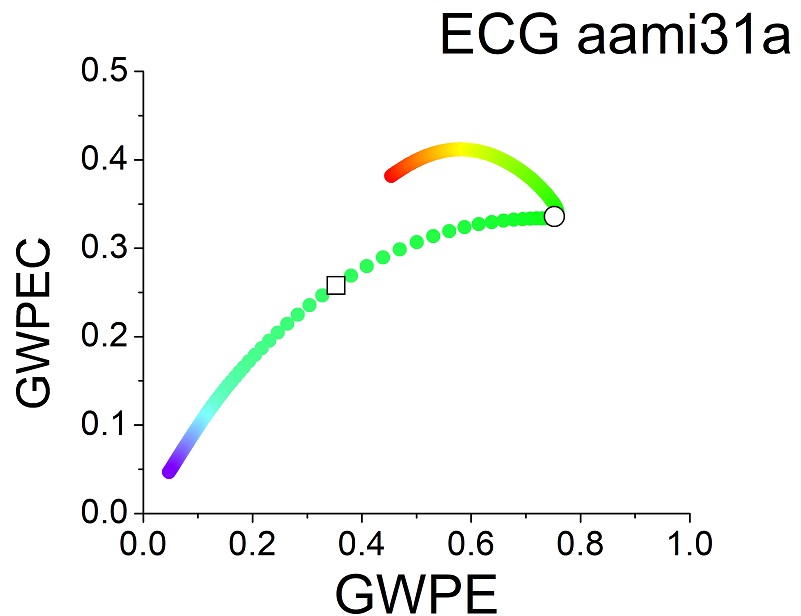}\\
\end{tabular}
\caption{
The trajectory of the points in the complexity-entropy causality plane with increasing $q$.
}
\label{fig2}
\end{figure}

Fig.~\ref{fig2} shows results of the calculations in the customary complexity-entropy causality plane representation \cite{rosso2007distinguishing}, where is seen that the stochastic series spectra display qualitatively similar behavior among themselves (with different placement of the PE and WPE points). Moreover, the three chaotic series are also rather similar among themselves, as well as the random series, while the ECG aami31a series demonstrates distinct behavior, perhaps the most reminiscent of fBm $H=0.9$.

\begin{figure}[h]
\begin{tabular}{ccc}
\includegraphics[width=.33\linewidth]{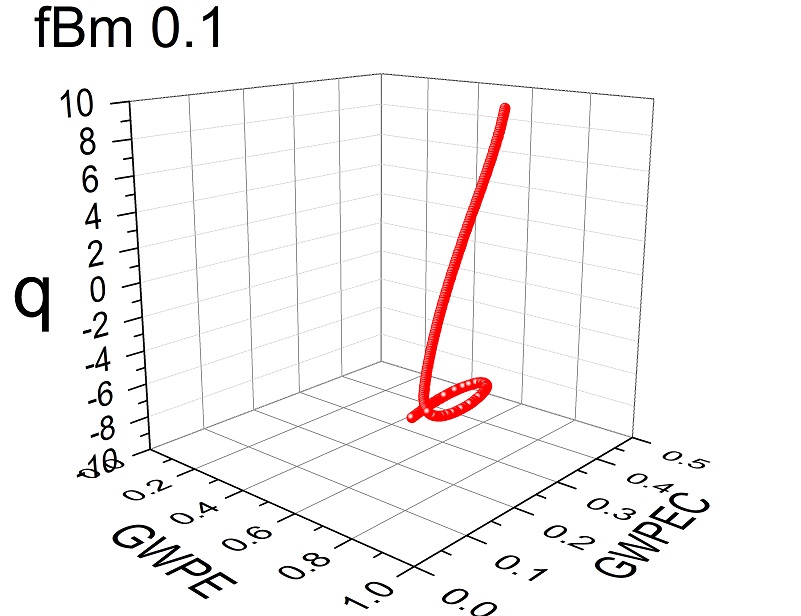} & 
\includegraphics[width=.33\linewidth]{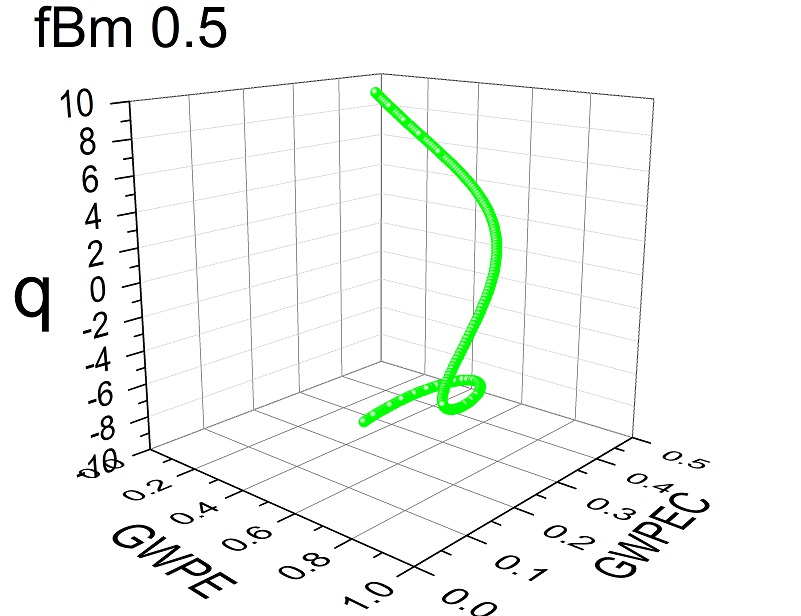} &
\includegraphics[width=.33\linewidth]{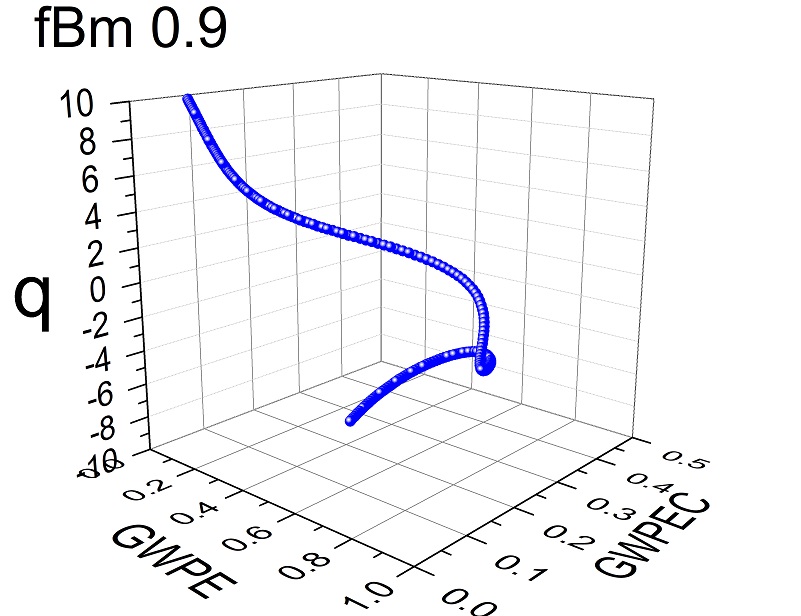}\\
\includegraphics[width=.33\linewidth]{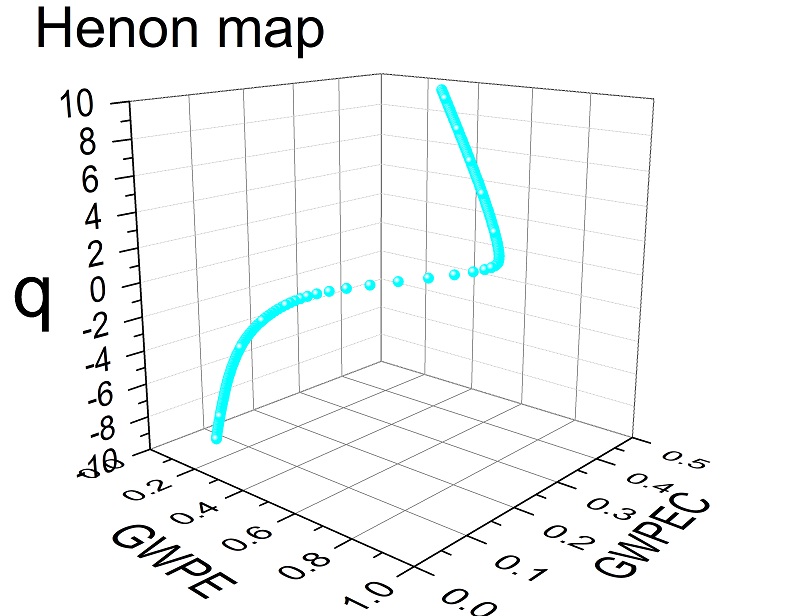} & 
\includegraphics[width=.33\linewidth]{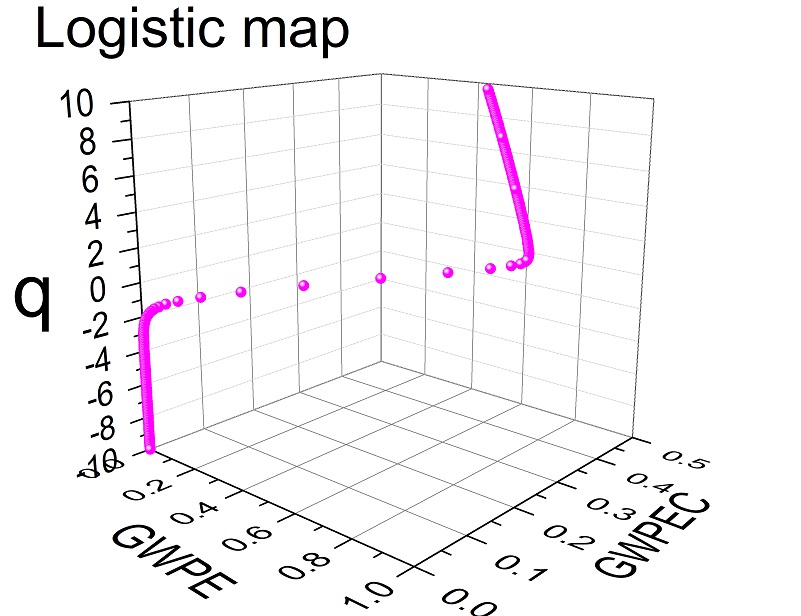} &
\includegraphics[width=.33\linewidth]{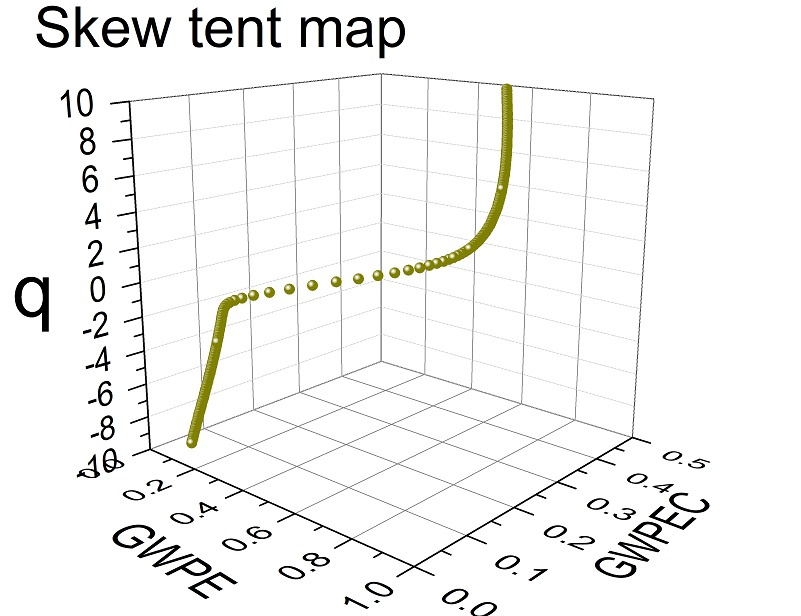}\\
\includegraphics[width=.33\linewidth]{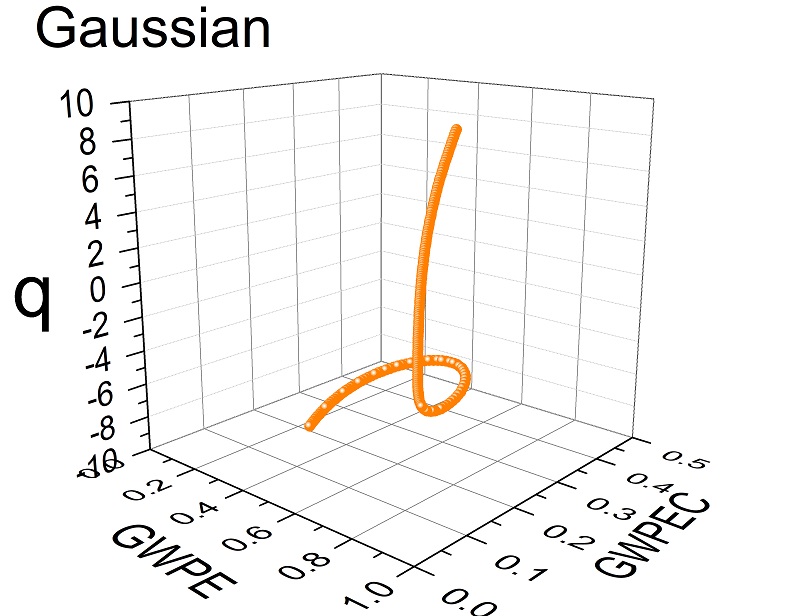} & 
\includegraphics[width=.33\linewidth]{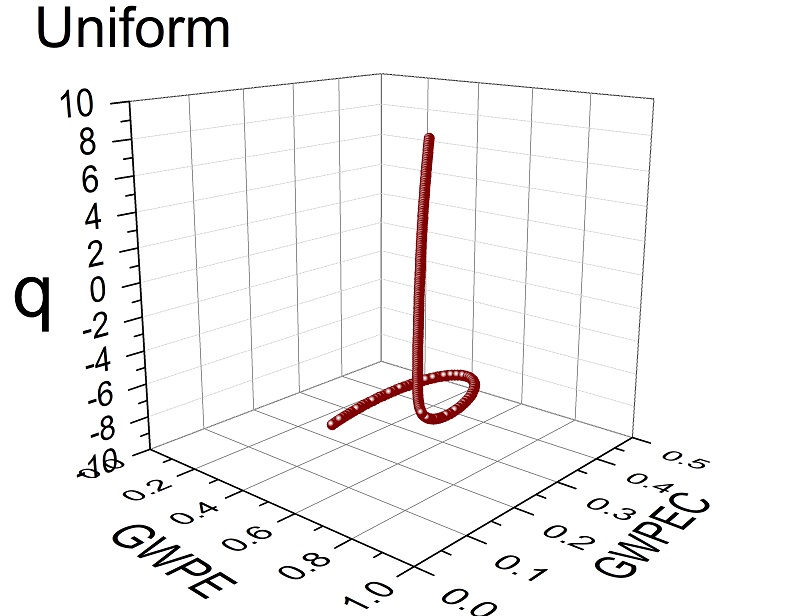} &
\includegraphics[width=.33\linewidth]{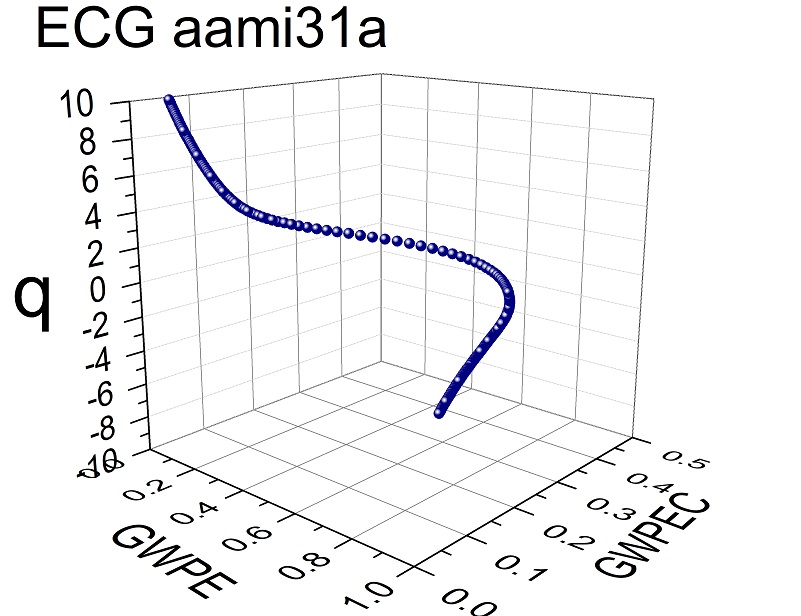}\\
\end{tabular}
\caption{
Signature curves of the considered samples in the complexity-entropy-scale causality box, with generalized weighted permutation entropy (GWPE), complexity (GWPEC), and magnification factor $q$, as coordinates.
}
\label{fig3}
\end{figure}

To further improve the distinction among the considered series,
the concept of complexity-entropy causality plane \cite{rosso2007distinguishing} can now be generalized to complexity-entropy-scale causality box (CESCB) representation, considering entropy, complexity, and magnification factor $q$ as coordinates in the three dimensional space. Each of the sequences may thus be represented by a characteristic signature curve in three dimensions, as shown in Fig.~\ref{fig3}. 
It is seen in Fig.~\ref{fig3} that the CESCB representation signature curves display rather different behavior, with a clear distinction between the stochastic process fBm series and the chaotic series. Moreover, there is a certain level of similarity between the shapes of the fBm series and the random series curves, the three chaotic processes curves have rather similar shapes, while the ECG series curve displays distinct behavior, with some similarity with the fBm $H=0.9$ series signature curve, particularly in the $q>0$ region.

\section{Conclusion}
In conclusion, the current study proposes a novel approach (GWPE) that sheds new light on the level of disorder of patterns of consecutive values, at different dispersion scales, combining and generalizing the PE and WPE methods, with the expectation that it may further enhance the numerous studies in areas where PE and WPE have already been successfully applied, as well as to make way for new applications with an improved understanding of the phenomenon at hand. The question may also be posed as to how the information from the signature curve shape may be synthesized into a single numerical quantifier that may serve for series classification in a rather general context, but the answer to this question requires further systematic studies in this direction, subject of current research to be published elsewhere, together with corresponding software and/or source code.

\section*{Software and Data Availability}
A self-contained C library with examples of implementation in C, R, and Python, and instructions for running examples addressed in this work in each environment (including data and results for word size $w=6$) is available on GitHub (https://github.com/stosicresearch/gwpentropy).

\section*{Acknowledgments}
T.S. acknowledges the support of Brazilian agency CNPq through grant 304497/2019-3.

\section*{References}
\bibliography{gwpentropy}

\end{document}